\documentstyle[preprint,aps]{revtex}
\begin{document}
\title{Chirally Extended QCD or An Asymptotically Free Chiral Linear Sigma
Model With Quarks Coupled to Gluons }
\author{Vikram Soni} \address{Theory group, National Physical Laboratory,\\ Dr.
K. S. Krishnan Road, New Delhi 110012, India.} \maketitle
\begin{abstract}
We show that the quark (not nucleon) based chiral linear $\sigma$ model which
receives strong support as a mean field theory for the strong interactions can
actually be made asymptotically free by coupling to gluons. It has the further
bonus of providing confinement.Alternatively, this theory is QCD that\ is
chirally extended by the addition of a $(\sigma,\pi)$ chiral multiplet. It
could therefore be a candidate theory for the strong interactions which is
somewhat different from QCD. \end{abstract}
\pacs{}

\section{Introduction}

Quantum Chromodynamics (QCD) is an acceptable starting point for a theory of
strong interactions. The main experimental support for this comes from the well
established phenomenon of asymptotic freedom which gives a qualitatively
correct description of deep-inelastic scattering (DIS) and scaling. However,
the increase of the coupling at low energy renders non-perturbative and
difficult the explicit low energy calculations.

The aspects of QCD that preserve at low energy are the symmetries. It is well
known that chiral symmetry is almost exact and furthermore, spontaneously
broken at low energy, a fact that actually precedes QCD. This has been given a
field theory realization via effective, but, renormalizable field theories like
the Gell-Mann Levy chiral $\sigma$ model with the pion as the goldstone boson.
Via PCAC this model provides us with various tested low energy theorems, for
example, the Goldberger-Treimann relation, the Adler-Weisberger relation etc.

The linear sigma model being a Yukawa theory is not asymptotically free, and
therefore, inspite of being pertubatively renormalizable , is usually
considered to be a low energy effective theory.Nevertheless, more recently, the
same chiral linear $\sigma$ model, with quarks~\cite{ref1} substituted for the
nucleons, has been found to provide a very reasonable description of the
nucleon~\cite{ref2} and strong interaction properties at finite temperature(
hadronic screening masses)~\cite{ref3,ref4} and baryon density~\cite{ref5},
even at scales well beyond chiral restoration (see also Ref.~\cite{ref6}).

This is unexpected from a theory that is only an effective theory. We shall
show in what follows that, surprisingly, the chiral linear $\sigma$ model with
quarks, when coupled to gluons can be asymptotically free.

\section{The Chiral Quark Sigma Model (with Gluons)}

The lagrangian is
\begin{eqnarray}{\cal L}=&-&\frac{1}{2}(\partial_\mu\sigma)^2-\frac{1}{2}(
\partial_\mu\vec\pi)^2-\lambda^2{(\sigma^2+\vec\pi^2-f_\pi^2)}^2\nonumber\\ &-&
\overline\Psi_q\left[{\cal D}_\mu+g_y(\sigma+i\gamma_5\vec\tau\cdot\vec\pi)
\right]\Psi_q-\frac{1}{4}G_{\mu\nu}^a G^{\mu\nu a}+{\cal O}(m\pi)\label{lagan}
\end{eqnarray}
where ${\cal D}_\mu=\partial_\mu-ig_3A_\mu^aT^a$ and $G_{\mu\nu}^a=\partial_\mu
A_\nu^a-\partial_\nu A_\mu^a+g_3f_{abc}A_\mu^b A_\nu^c$ \\$A_\mu^a$ is the
color gluon field and $T^a$ is the SU(3) color matrix in the fundamental
representation. $g_y$, $g_3$ and $\lambda$ are the Yukawa, QCD, and Higgs
couplings respectively. $\overline\Psi_q$ is the quark field.

Apart from the quarks and the chiral fields ($\sigma,\pi$) we have additionally
included the gluon interaction with the quarks. Note that the chiral fields ($
\sigma,\pi$) are color singlets and do not interact with the gluons. The
relevance of these gluons will be clarified later. Note that this lagrangian is
alternatively nothing but QCD extended by the addition of a $(\sigma,\pi)$
chiral multiplet. First let us examine the support this lagrangian receives at
the mean field level from very different quarters, even in the neglect of
gluons.

\section{Mean Field Results} \subsection{Lattice Results}

Recently, lattice QCD results~\cite{ref3,ref4} which look at the chiral
transition at finite T, in the presence of dynamical quarks, give an
interesting picture which we shall briefly recount.

We focus on the screening mass spectrum for the hadrons as the chiral
restoration temperature, $T_\chi$, is approached. The menagerie of hadrons
(like the nucleon, the $\rho$, the $A_1$ etc.) begin to scale as their free
constituent quarks and antiquarks. For example, for $T>T_\chi$ the nucleon
screening mass is just three times the screening mass of the free quark,
whereas the $\rho$ and $A_1$ screening masses are two times the free
quark(antiquark) screening mass.
\begin{equation}\frac{M_N}{M_\rho}\sim\frac{3}{2}\hspace{2em}{\rm and}\hspace{
3em}\frac{M_{A_1}}{M_\rho}\sim1\end{equation}

However, the pion and sigma screening masses, though they are degenerate, at $T
_\chi$ (chiral restoration), stay well below the free q\={q} threshold all the
way till about $3T_\chi$, the maximum temperature achieved by the lattice
calculations
\begin{equation}\frac{M_{\pi,\sigma}}{M_{\rho,A_1}}<1\end{equation}
Thus, the pion and sigma continue to be bound states all the way till the
maximum temperature reached on the lattice. Calculations for larger $T$ will
yield even more information.

Interestingly enough, as indicated by Goksch~\cite{ref3}, a finite temperature
version of the chiral $\sigma$ model with quarks instead of nucleons can
reproduce the lattice results rather well. Of course, in this instance the
pions (sigma) are elementary and will continue to exist for all $T>T_\chi$,
except not as Goldstone bosons. Whether such a scenario will match lattice
results at higher temperature is not known. Nevertheless it shows that the
chiral $\sigma$ model with quarks is an effective lagrangian well above the
chiral restoration temperature. Gluons may be included perturbatively but have
been neglected in Ref.~\cite{ref3}.

\subsection{The Nucleon}

In Ref.~\cite{ref1} it was found that in the spontaneously broken phase this
lagrangian yields the nucleon (ground state in the $B=1$ sector) in a Skyrme
background with bound state quarks. Nucleon static properties are reasonably
described at the mean-field level~\cite{ref2}

Further, the nucleon realised (as above) as a chiral quark soliton in a Skyrme
background is (like the Skyrmion) roughly consistent with the EMC result, that
the axial vector direct form factor (which is twice the total `spin' of the
quarks) is consistent with zero for the proton state~\cite{ref6,ref7,ref8}.
This is in contrast to the non-relativistic quark model for which the axial
vector direct form factor is obviously close to unity.

\subsection{Equation of State} A very interesting support for the lagrangian
over the entire range of bayron density was demostrated by one of
us~\cite{ref5} in Mean Field Theory (MFT).

It was shown in Ref.~\cite{ref5} that the model lagrangian has two phases at
the MFT level. Starting from zero density we get an equation of state where the
ground state occurs in phase~1 -- solitonic nucleons composed of bound quarks
in a skyrme background, with a qualitatively correct binding energy and an
absolute minimum in $E_B$ at around, $\rho_{\rm nuc}$. The phase 2 in this
model is the so called Abnormal or Lee-Wick phase in which the condensate (VEV)
is space uniform. This provides a Yukawa or spontaneous mass for th quark. The
VEV is a function of baryon density going to zero at some critical density.At
density higher than this we go into a chirally restored phase. The transition
from phase 1 to phase 2 occurs at a density when the latter is well into chiral
restoation and the ground state is described by a free fermi gas of quarks. The
two phases thus conspire to recover the expected features of the strong
interaction equation equation of state at all densities

The results quoted here are indicative of the fact that Eq.~(\ref{lagan}) is a
good effective lagrangian at scales even well above chiral restoration. Is it
possible, then, that this theory may be more than just an effective theory. In
this case, it would need to be asymptotically free, and to this we now turn.

\section{Beyond Mean Field Theory, Gluons and Asymptotic freedom}

We now come to a feature which has to do with the lack of asymptotic freedom in
simple yukawa theories~\cite{ref9} like the chiral $\sigma$ model. In the
attempt to go beyond MFT to 1-loop order one encounters the vacuum instability
to small length scale fluctuations~\cite{ref10} that occurs for all simple
yukawa theories.

This is related to the fact that due to the lack of asymptotic freedom the
yukawa coupling has a Landau singularity~\cite{ref11}. This fact does not
permit us to go beyond the MFT level in an otherwise (perturbatively)
renormalizable theory~\cite{ref12}. It has been observed that the wave function
renormalization for the scalar(psuedoscalar) is inversely proportional to the
running yukawa coupling~\cite{ref13}. This implies that when the yukawa
coupling blows up, the wavefunction renormalization, ${\cal Z}$, goes to
zero.\footnote{We shall prove this shortly via the RNG} The scale at which this
happens is often interpreted as a compositeness scale for the scalar i.e., at
and after this scale the scalar ceases to be (elementary).

In an attempt to circumvent this problem we had proposed the inclusion of
gluons that interact directly only with the quarks~\cite{ref6}. Being vector
bosons they clearly stem the rise of the yukawa coupling as can be seen from
the following calculation. Further, the gluons are important for confinement, a
feature that is absent in the chiral quark $\sigma$ model.

The significant question then is : Can the theory described by Eq.(1) possibly
be asymptotically free. We now turn to this.

This question can be answered by looking at the $\beta$-function for the yukawa
coupling and the QCD coupling for our lagrangian given in Eq.~(\ref{lagan}).

The $\beta$ function for the QCD coupling, $\alpha$, is
\begin{equation}\frac{\partial\alpha}{\partial t}=-\left(\frac{33-2N_F}{3}
\right )\frac{\alpha^2}{8\pi^2}\hspace{5em}\left(g_3^2=\alpha\right )\label
{dadt}\end{equation}
This is for $m_q=0$ and $t=\ln(p/\mu)$.\\ Clearly this is valid for small
$\alpha$, or only in the ultraviolet when we can take $N_G=3$ or $N_F=6$
(Remember all quarks couple to the gluons). Note, that to one-loop order the
$\beta$-function for the QCD coupling does not receive any contribution from
the yukawa coupling, $g_y$, or the scalar self coupling $\lambda$, as the
scalar (pseudoscalars) do not couple directly to the gluons.

The yukawa coupling $g_y$ for the pion and sigma to the quarks has the
following $\beta$ function (Remember the pion and sigma couple to only one, the
[u,d], generation. We have assumed that the $\pi$ and $\sigma$ coupling to the
other generations is absent).
\begin{equation}\frac{\partial{g_y^2}}{\partial t}=\frac{g_y^2}{8\pi^2}\left[12
g_y^2-8\alpha\right]\label{dgy}\end{equation}
One can see here explicitly how the QCD coupling, $\alpha$, slows down the
increase of the yukawa coupling, $g_y^2$, with $t$.

We can now define the ratio $\rho=g_y^2/\alpha$ and write the following
equation for $\rho$ using Eqs~(\ref{dadt}) and (\ref{dgy}). This has the
advantage that $\rho$ can be expressed as a function of a single variable $
\alpha$
\begin{equation}\frac{\partial\rho}{\partial{\alpha}}=-\frac{\rho}{\alpha A}[12
\rho-8+A]\label{DrByDa}\end{equation}
where $A=(33-2N_F)/3$ and $N_F$ is the number of flavours that effectively
couple to the gluons. For the case of 3 generations or $N_F=6$, (This will
correspond to the high energy behaviour of this theory).
\begin{equation}\frac{\partial\rho}{\partial{\alpha}}=-\frac{\rho}{7\alpha}[12
\rho-1]\end{equation}
Here we observe that there are two regimes. \begin{itemize} \begin{description}
\item[(i) $0<\rho<1/12$] \hfill

In this case $\partial\rho/\partial\alpha>0$. This implies that $\rho$
decreases as $\alpha$ decreases, that is, $\rho$ will decrease with increasing
momentum scale. We can integrate the $\rho$ equation to get
\begin{equation}\rho=\frac{\alpha^{1/7}K}{(1+12\alpha^{1/7}K)}\end{equation}
where $K$ is a positive integration constant that is set by initial data on $
\alpha$ and $g_y^2$ This equation is interesting. It shows that in the
ultraviolet when $\alpha\rightarrow0$
$$\rho\sim K\alpha^{1/7}$$
This further implies that in the ultra violet
$$g_y^2\sim K\alpha^{8/7}$$
which shows that $g_y^2$ is asymptotically free and goes to zero faster than
$\alpha$. Therefore, the leading behaviour of this theory in the ultraviolet is
given by the QCD coupling with the yukawa coupling contributing only in sub
leading order.

On the other hand if we go to $\alpha\rightarrow\infty$ we find $\rho
\rightarrow1/12$. Since, in the 1-loop approximation, $\alpha\rightarrow\infty$
at some length scale, $\Lambda$, we may want to think of $\rho=1/12$ as a quasi
fixed point~\cite{ref14,ref15} (fixed point in the ratio of the coupling) in
the infrared. Evidently, such an interpretation is fraught with uncertainty as
the 1-loop analysis breaks down well before $\alpha\rightarrow\infty$. Thus the
quasi fixed point $\rho=1/12$, cannot be taken seriously unlike the
Ross-Pendelton fixed point~\cite{ref14} for the standard electroweak model
where both $\alpha$ and $g_y^2$ are small in the infrared (250~GeV) and the
quasi fixed point is meaningful.

\item[(ii) $\rho>1/12$] \hfill

Here ${\partial\rho}/{\partial{\alpha}}<0$ and therefore $\rho$ increases as $
\alpha$ decreases that is $\rho$ increases with the momentum scale. The
integration of the $\rho$ equation now gives
\begin{equation}\rho=\frac{\alpha^{1/7}}{(12\alpha^{1/7}-C)}\label{rhoeq}
\end{equation}
where $C$ is a positive constant set by data on $\alpha$ and $g_y^2$.

However, this equation is very different. Naively, when $\alpha\rightarrow
\infty$ we recover the so called quasi fixed point, $\rho=1/12$. But as we have
pointed out earlier this quasi fixed point is not really meaningful. On the
other hand as we move towards the ultraviolet and $\alpha$ comes down, there is
a pole in the denominator at $C=12\alpha^{1/7}$ indicating that $\rho$ blows
up, which in turn means that the yukawa coupling $g_y^2$ blows up even as $
\alpha$ is finite. The scale at which this happens, as we shall see shortly,
can be identified with the vanishing of the wavefunction renormalization for
the scalar(pseudoscalar) where the scalar(pseudoscalar) is no longer an
elementary particle.

The regime can therefore be thought of as a phase in which the
scalar(pseudoscalar)is composite. ( A complete investigation fo these results
which includes the scalar self coupling will be reported
separately~\cite{ref16,ref17}).

\end{description} \end{itemize}

\section{The wavefunction Renormalization for the Scalar(Psuedoscalar)}

The wavefunction renormalization for the scalar( psuedoscalar) going to zero
has the interpretation of the particle being composite i.e., not elementary,
whereas if it does not go to zero the particle is elementary. We shall now
calculate the renormalization group (RNG) improved wavefunction renormalization
to show its relation to the yukawa coupling in both the regimes (i) $0<\rho<1/
12$ and (ii) $\rho>1/12$ .

Let us now briefly consider the wavefunction renormalization for the scalar
(psuedoscalar) fields to see how it relates to the yukawa and QCD couplings. To
bring out the right connection we must carry out a RNG improvement of the two
point function. The function can be calculated in perturbation theory.
\begin{equation}{\cal Z}^{(2)}_{\rm1-loop}=1+2B g_y^2t\hspace{5em}(m_q=0)
\end{equation}
where $B=-4N_c/16\pi^2$ and $t=\log(p/\mu)$.

The $\gamma$ function is $\gamma=B g_y^2$. The RNG improvement proceeds as
follows
\begin{equation}\left(-\frac{\partial}{\partial t}+\beta\frac{\partial}{
\partial g_y}+2\gamma_0\right ){\cal Z}^{(2)}=0\hspace{4em}{\rm where}\hspace{
1em}\gamma_0=\gamma(t=0)\end{equation}
The RNG improved ${\cal Z}^{(2)}$ is then
\begin{equation}{\cal Z}^{(2)}=\exp2\int_0^t\gamma(t') dt'\end{equation}
Using the two equations of the $\beta$ functions for the QCD and yukawa
coupling, we can cast the $\gamma$ function
\begin{equation}\gamma(g_y^2(t'))=B g_y^2(t')=\frac{B16\pi^2}{4N_c}\frac{
\partial\log(g_y)}{\partial t'}-\frac{8B}{4N_c}\frac{16\pi^2}{A}\frac{\partial
\log(g_3)}{\partial t'}\end{equation}
It follows
\begin{equation}{\cal Z}^{(2)}={\left(\frac{g_y(t)}{g_y(t=0)}\right )}^{-2}
\left(\frac{g_3(t)}{g_3(t=0)}\right )^{16/7}={\left(\frac{g_y(t)}{g_y(t=0)}
\right )}^{-2}\left(\frac{\alpha(t)}{\alpha(t=0)}\right )^{8/7}\end{equation}
This is a proof for the inverse relation between the yukawa coupling and the
wavefunction renormalization ${\cal Z}^{(2)}$.

Furthermore, it is interesting to see what happens to ${\cal Z}^{(2)}$ in the
ultraviolet for \begin{itemize} \begin{description} \item[(i) $0<\rho<1/12$:]
\hfill

Here we know $g_y^2\approx K(\alpha)^{8/7}$ in the ultraviolet. This gives ${
\cal Z}^{(2)}=1$ a rather interesting result. \item[(ii) $\rho>1/12$:] \hfill

$g_y(t)\left|_{\rm uv}\rightarrow\infty\right .$ with ${\cal Z}^{(2)}
\rightarrow0$ in the ultraviolet \end{description} \end{itemize}

If we find ourselves in the regime $\rho<1/12$ where the yukawa coupling as
asymptotically free there is a rather unorthodox consequence. The pion
continues to be elementary to all ultraviolet scales. This is strange but
lattice gauge theories do indicate an elementary non-dissociated-pion (sigma)
till $T\approx3T_\chi$. Better lattice gauge theory calculations could test
this conjecture, which will have interesting physical consequences.

\section{Remarks} \begin{itemize} \begin{enumerate} \item We would like to
emphasize that our results depend importantly on our taking the number of
flavours, $N_F=6$ (or number of generations=3) for QCD $\beta$ function. This
is true for the scales in the ultraviolet ($q^2\gg m_t^2$) but not in the
infrared. Anyhow, we do not trust the 1-loop results in the infrared.

However if we do take the number of flavours to be $N_F=2$ (just u and d
quarks) then the value of A in the Eq.~(\ref{DrByDa}) becomes $29/3$ and
\begin{equation}\frac{\partial\rho}{\partial\alpha}=\frac{\rho}{\alpha}\frac{3}
{29}\left[\frac{1}{12\rho}+\frac{5}{3}\right]\end{equation}
This moves the asymptotically free regime to negative $\rho$ and there is no
ready interpretation for this.

\item We have not addressed the question of the asymptotic freedom of the
scalar(pseudoscalar) self coupling, $\lambda$. However, there is an analysis by
Schrempp and Schrempp~\cite{ref15} for the standard model which deals with the
two ratios of the coupling, $\rho=g_y^2/\alpha$ and $R=\lambda/g_y^2$ that can
be adapted to our model. This analysis is carried out in
Refs~\cite{ref16,ref17}. We make only some brief observations from these
results here. The $\beta$-function for $\lambda$ in our model is
\begin{equation}\frac{\partial\lambda}{\partial t}=\frac{1}{8\pi^2}\left[2
\lambda^2-144g_y^4+24g_y^2\lambda\right]\end{equation}
Again, by defining the ratio $R=\lambda/g_y^2$ we can convert to an equation
for $R$ that depends on the single variable $\rho$ and find that
\begin{equation}\frac{\partial R}{\partial\rho}=\frac{1}{[12\rho-8+A]}\left[2R^
2+R\left(12+\frac{8}{\rho}\right)-144\right]\end{equation}

It is found that at least on a single trajectory in the $[R,\rho]$ parameter
space, that is the invariant line~\cite{ref15,ref16,ref17} the ultra violet
behaviour of $R$ for the regime $\rho<1/12$ is such that $R\rightarrow0$ in the
ultraviolet. This would indicate that our results are also intact insofar as
the coupling $\lambda$ is concerned. $\lambda\rightarrow0$ in the extreme
ultraviolet even faster than $g_y^2$. As we have indicated these results will
be reported separately where we shall present a complete analysis of such field
theories~\cite{ref16,ref17} and where we will also look at tests to decide on
any of these theories can compete with QCD as far as the physics of the
ultraviolet (DIS) is concerned~\cite{ref17} \end{enumerate} \end{itemize}

\section{Conclusion} \begin{itemize} \item We have shown the existence of an
asymptotically free chiral theory of quarks mesons and gluons with yukawa and
gluon couplings such that the leading ultraviolet behaviour, like QCD, comes
from the gluon coupling and only subleading effects arise from the yukawa
coupling. The 1-loop calculation indicates this is true for the `phase' of the
theory that has $0<\rho<1/12$. In other words it depends on a set of initial
data on the QCD and yukawa couplings of the theory. Note, also that this
initial data must be gleaned (experimentally) from the ultraviolet. Data for
the infrared cannot really be used as our 1-loop calculation breaks down for
large coupling. \item This theory implies an elementary pion for all scales
\item For initial data, when $\rho>1/12$, in our 1-loop approximation (note,
this number will change if we go beyond 1-loop ! ) `there is another phase' of
the theory. In this `composite' phase, the yukawa coupling acquires a Landau
singularity and the theory will lose its elementary scalar (pseudosclar). \item
Such a theory has infrared quasi fixed points (fixed point in the ratio of the
couplings). \end{itemize}

We have shown that our theory is asymptotically free and thus a candidate for a
consistent theory of the strong interactions valid at all scales. This begs a
new question how is this theory different from QCD ?

The quark based chiral $\sigma$ model receives strong support at the mean field
level as a candidate to describe the strong interaction. We have shown that by
coupling to gluons it can be made asymptotically free depending on initial data
for the QCD, yukawa and meson self coupling. A bonus is that it can then also
provide confinement. We strongly believe it needs further study.

\section{Acknowledgements}

Much of this work came out of discussions with Bachir Moussallam at IPN Orsay .
The author would like to thank him for his dogged insistence, that lack of
asymptotic freedom in the chiral $\sigma$ model must be rectified before we can
make further progress. The author would also like to thank P. S. Discussions
particularly with N. D. Haridass and with S. H. Sharat Chandra, R. Anishetty
and R. Basu are acknowledged. This work has a strong link with my father and I
would like to share it with him. I would like to thank Kapil Bajaj for ty(p)ing
up the `loose ends'

\end{document}